\documentclass[twocolumn,superscriptaddress,showpacs,letterpaper]{revtex4-1}%
\makeatletter

\newcommand{\Rmnum}[1]{\expandafter\@slowromancap\romannumeral #1@}
\makeatother
\usepackage{graphicx}
\usepackage{sidecap}
\usepackage{latexsym}
\usepackage{amssymb}
\usepackage{amsfonts}
\usepackage{soul}
\usepackage{color}
\usepackage{amsmath}
\usepackage{geometry}
\usepackage[colorlinks,linkcolor=blue, urlcolor=blue, anchorcolor=blue, citecolor=blue]{hyperref}

\begin{document}

\title{Suppression of flux jumps in high-$J_c$ Nb$_3$Sn conductors by ferromagnetic layer}

\author{Cun Xue}
\email{xuecun@nwpu.edu.cn}
\affiliation{School of Mechanics, Civil Engineering and Architecture, Northwestern Polytechnical University, Xi'an 710072, China}

\author{Kai-Wei Cao}
\affiliation{School of Mechanics, Civil Engineering and Architecture, Northwestern Polytechnical University, Xi'an 710072, China}

\author{Tian He}
\affiliation{Materials Genome Institute, Shanghai University, Shanghai 200444, China}

\author{Chong Wei}
\affiliation{School of Mechanics, Civil Engineering and Architecture, Northwestern Polytechnical University, Xi'an 710072, China}

\author{Wei Liu}
\email{liuwei17@c-wst.com}
\affiliation{Western Superconducting Technologies Co., Ltd., and Xi'an Superconducting Magnet Technology Co., Ltd, Xi'an 710014, China}

\author{Jun-Yi Ge}
\email{junyi_ge@t.shu.edu.cn}
\affiliation{Materials Genome Institute, Shanghai University, Shanghai 200444, China}

\date{\today}

\begin{abstract}

Flux jumps observed in high-$J_c$ Nb$_3$Sn conductors are urgent problems to construct high field superconducting magnets. The low-field instabilities usually reduce the current-carrying capability and thus cause the premature quench of Nb$_3$Sn coils  at low magnetic field. In this paper, we explore suppressing the flux jumps by ferromagnetic (FM) layer. Firstly, we experimentally and theoretically investigate the flux jumps of Nb$_3$Sn/FM hybrid wires exposed to a magnetic field loop with constant sweeping rate. Comparing with bare Nb$_3$Sn and Nb$_3$Sn/Cu wires, we reveal two underlying mechanisms that the suppression of flux jumps is mainly attributed to the thermal effect of FM layer for the case of lower sweeping rate, whereas both thermal and electromagnetic effects play a crucial role for the case of higher sweeping rate. Furthermore, we explore the flux jumps of Nb$_3$Sn/FM hybrid wires exposed to AC magnetic fields with amplitude $B_{a0}$ and frequency $f$. We build up the phase diagrams of flux jumps in the plane $f$-$B_{a0}$ for bare Nb$_{3}$Sn wire, Nb$_{3}$Sn/Cu wire and Nb$_{3}$Sn/FM wire, respectively. We stress that the region of flux jumps of Nb$_{3}$Sn/FM wire is much smaller than the other two wires, which indicates that the Nb$_{3}$Sn/FM wire has significant advantage over merely increasing the heat capacity.  The findings shed light on suppression of the flux jumps by utilizing FM materials, which is useful for developing new type of high-$J_c$ Nb$_{3}$Sn conductors.

\end{abstract}

\maketitle
%ratchet11

\section{introduction}
With high critical current density (non-matrix $J_c$ exceeds 3000 A/mm$^{-2}$ at 4.2 K, 12 T) and upper critical field (28-30 T) \cite{A1,A2,A3,A4}, Nb$_3$Sn conductor is a preferred material to construct high magnetic field (10-16 T) that is beyond the limit of NbTi \cite{A5}. Additionally, a hybrid solution of high temperature superconductor (HTS) and Nb$_3$Sn materials is usually used to achieve much higher magnetic field ($>$ 20 T) in order to avoid significantly high cost of HTS \cite{A6}. However, Nb$_3$Sn magnets are susceptible to the flux jumps  \cite{A7,A8,A9}, which seriously limits the Nb$_3$Sn performance. After magnetic flux jumps were first observed and investigated in 1960s \cite{A10}, the underlying physical mechanisms \cite{A11,A12,A13,A14} and the relevant physical parameters controlling the thermomagnetic instability (temperature \cite{A15}, ramping rate \cite{A16}, sample size \cite{A17}, border defects \cite{A18}) were gradually revealed. For composite superconducting wires, the early criterion for triggering the magnetic flux jump was proposed by Swartz \& Bean \cite{A19} and Wilson \cite{A20}. Subsequently, in order to further improve the performance of Nb$_3$Sn high-field magnets, researchers conducted a series of experimental studies \cite{A21,A22,A23} and theoretical analyses \cite{A24,A25,A26} to describe the characteristics of low-field flux jumps. During the flux jumps, substantial energy is released in the form of Joule heating within a very short time due to the rapid motion of magnetic flux trapped inside the filaments. In this case, the flux instability behaviors usually lead to undesirable problem of premature quenches \cite{A27}.Experimental results demonstrated that the superconducting wires that carry high currents at high fields (greater than 10 T) cannot sustain these same currents at low fields (1–3 T) when the sample current is fixed and the magnetic field is ramped \cite{A28}. Moreover, frequent and sudden spikes in voltage signals sometimes are very high, which can cause difficulties for the digital quench detection system \cite{A29}.

In recent years, researchers have shown more and more academic interests in exploring effective methods to prevent the flux instabilities of Nb$_3$Sn. Experiments by Fermi national accelerator laboratory \cite{B1,B2} and Brookhaven national laboratory \cite{B3,B4} show that both reducing the filament size ($d_{eff}$) and increasing residual resistivity ratio (RRR) of the copper can significantly improve stability of Nb$_3$Sn at low field, which can indeed improve Nb$_3$Sn conductor performance and current carry ability. On one hand, more subelements can reduce $d_{eff}$. On the other hand, however, it may also easily cause the problems of contamination and thus raise manufacturing difficulties \cite{B5}. The higher RRR can be achieved by reducing the reaction time or by modifying the design with adjusting the metal ratio of Nb and Sn, which is at the cost of reducing the critical current density \cite{B6}. Wilson et al demonstrated that the intrinsic stability of high-$J_c$ Nb$_3$Sn conductors can be improved by adjusting the temperature margin and the dynamic cooling \cite{A21}. However, further increasing the temperature margin beyond a certain limit becomes increasingly challenging for Nb$_3$Sn utilized in generating high fields. Condisering Nb$_3$Sn coils are usually impregnated with epoxy, the quite small thermal diffusion coefficient may weaken the effectiveness of dynamic cooling. The experiments by Goldfarb et al \cite{B8} showed that the number of flux jumps decreases remarkably when the high-$J_c$ Nb$_3$Sn wire exposed to liquim helium at 2.1 K, which indicates that improvement of heat transfer capacity indeed can suppress the low-field instabilities. This method needs very costly lower operation temperature. Additionally, it may not suitable for Nb$_3$Sn magnets with densely packed coils where heat dissipation conditions of inner Nb$_3$Sn wires is completely different from the sample of single Nb$_3$Sn wire in experiments. Very recently, Xu et al \cite{B9} investigated the influences of heat treatment temperature and Ti-dopping on the low-field flux jumps. It was found that the suppression of low-field flux jumps by heat treatment temperature is attributed to both reducing the $J_c(B)$ curve slope and increasing the heat capacity. The negative result is a significant reduction of $J_c$ at low fields. Another promising idea to improve the thermomagnetic stability stems from increasing the specific heat. Xu et al proposed to introduce high specific heat substances (Gd$_2$O$_3$) in certain positions \cite{B10}. The experiments show that although many small flux jumps are still observed at low fields the amplitude of flux jumps can be indeed suppressed significantly. Therefore, it is still desirable to explore new methods to suppress the flux jumps.

It is well-known that the superconductors and ferromagnetic (FM) materials exhibit completely different electromagnetic behaviors. When coupling the two materials, it was expected to observe interesting phenomena in hybrid superconductors/ferromagnetic (SC/FM) structures. Many researchers developed theoretical and experimental approaches to explore the impacts of ferromagnetic substrate on the electromagnetic responses of superconductors. The complex-field approach with conformal mapping was used to analyze the effect of soft ferromagnetic substrate on the electromagnetic response of superconducting strips \cite{F1} and ac loss of 2G HTS power transmission cables  \cite{F2}. Recently, Prigozhin et al proposed a thin shell model and numerical scheme of Chebyshev spectral method for electromagnetic response of a coated conductor with magnetic substrate to variations of the transport current and applied field. The influence of a magnetic substrate on the superconducting current and ac losses is investigated\cite{F3}. Silhanek et al developed a quantitative magneto-optical imaging technique to directly visualize the thermomagnetic instabilities of superconductor/ferromagnet hybrid structures \cite{F4}. Another fascinating phenomenon is the so-called magnetic cloak to exactly cloak uniform static magnetic fields by specially designed superconductor-ferromagnetic structures \cite{F5,F6}. Due to the electromagnetic interactions, it is expected that the FM materials have clear impact on the low-field instabilities of Nb$_3$Sn wires, which has not been well explored yet.

In this paper, we explore the thermomagnetic instabilities of Nb$_3$Sn/ferromagnetic (Nb$_3$Sn/FM) hybrid wires exposed to time-varying magnetic field. The magnetization measurements show that the flux jumps of Nb$_3$Sn wires can be significantly suppressed by the FM layer. Combining with numerical simulations, we reveal different underlying mechanisms for suppression of flux jumps in the cases of low and high magnetic field sweeping rates, respectively. Then we further explore the flux jumps of Nb$_3$Sn/FM hybrid wire exposed to the AC magnetic field. We demonstrate that the flux jumps depends on both amplitude $B_{a0}$ and frequency $f$ of AC magnetic field. The phase diagram of flux jumps in the plane $f$-$B_{a0}$ for bare Nb$_{3}$Sn wire, Nb$_{3}$Sn/cooper (Nb$_{3}$Sn/Cu) wire and Nb$_{3}$Sn/FM wire are presented. The paper is organised as follows: we present the experimental measurements and theoretical method that are used to study the flux jumps of Nb$_3$Sn/FM wires in Section II. Then we discuss the suppression of flux jumps for two cases of a linear ramping magnetic field and AC magnetic filed. The summary is given in the final section.

\begin{figure}[!t]
\centering
\includegraphics*[width=3.3in,keepaspectratio]{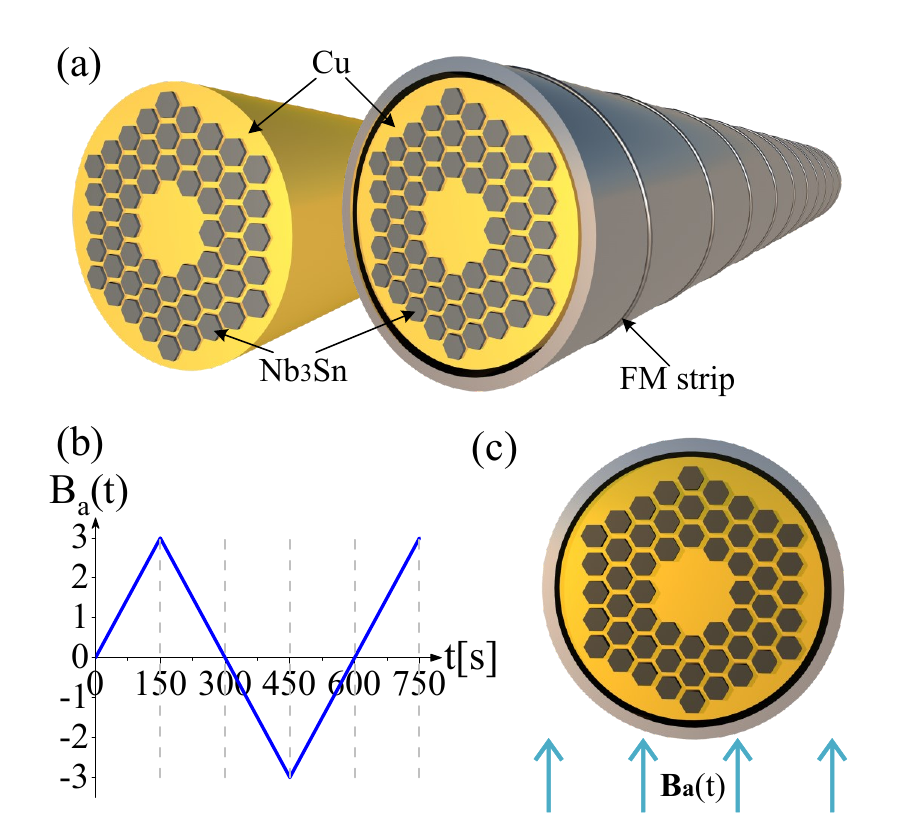}
\caption{(a) Bare Nb$_{3}$Sn wire with 54 subelements (left) and hybrid-structured Nb$_{3}$Sn wire wrapped by FM strip (Nb$_3$Sn/FM wire, right). (b-c) The samples are exposed to an applied magnetic field with sweeping rate of 0.02 T/s.}
\end{figure}

\section{Experiments setup and THEORETICAL FORMALISM}
To prepare the samples used in the experimental measurements, we choose the multi-filamentary Nb$_3$Sn wire manufactured with the so-called internal-tin approach. The diameter of Nb$_3$Sn wire is about 1.3 mm and its subelement structure is shown in Fig. 1(a). In this work, two kinds of samples are fabricated, i.e., bare Nb$_3$Sn wire (see left panel in Fig. 1{(}a{)}) and Nb$_3$Sn wire warped by a FM layer made of 1J22 soft magnetic alloy (right panel in Fig. 1{(}a{)}). The data of the samples are shown in the Appendix. We prepare three samples of hybrid Nb$_3$Sn/FM wires and the thickness of the FM layer is 0.1 mm, 0.2mm and 0.3 mm, respectively. A polishing procedure is performed for both ends of all samples. In real engineering applications, the Nb$_3$Sn wire is generally exposed to an external magnetic field and transport current with ramping rates normally less than 5A/s. Actually, simulation results show that the effect of transport current on number of flux jumps and amplitude of flux jumps is very slight when ramping rates of transport current is less than 5 A/s \cite{XS1}. Therefore, in order to detect the thermomagnetic instabilities, we perform the magnetization measurements for all samples only exposed to a transverse magnetic field loop with Magnetic Property Measurement System (MPMS) at $T_0=4.2$ K. Note that a zero-field cooling down from 20 K to 4.2 K is first performed before each magnetization measurements. As shown in Fig. 1(b), the magnetic field is varying between +3 T and -3 T with a constant field-sweeping rate of 20 mT/s, i.e., $B_a=$ 0 T $\rightarrow$ 3 T $\rightarrow$ -3 T $\rightarrow$ 3 T.

\begin{figure}[!t]
\centering
\includegraphics*[width=0.9\linewidth,angle=0]{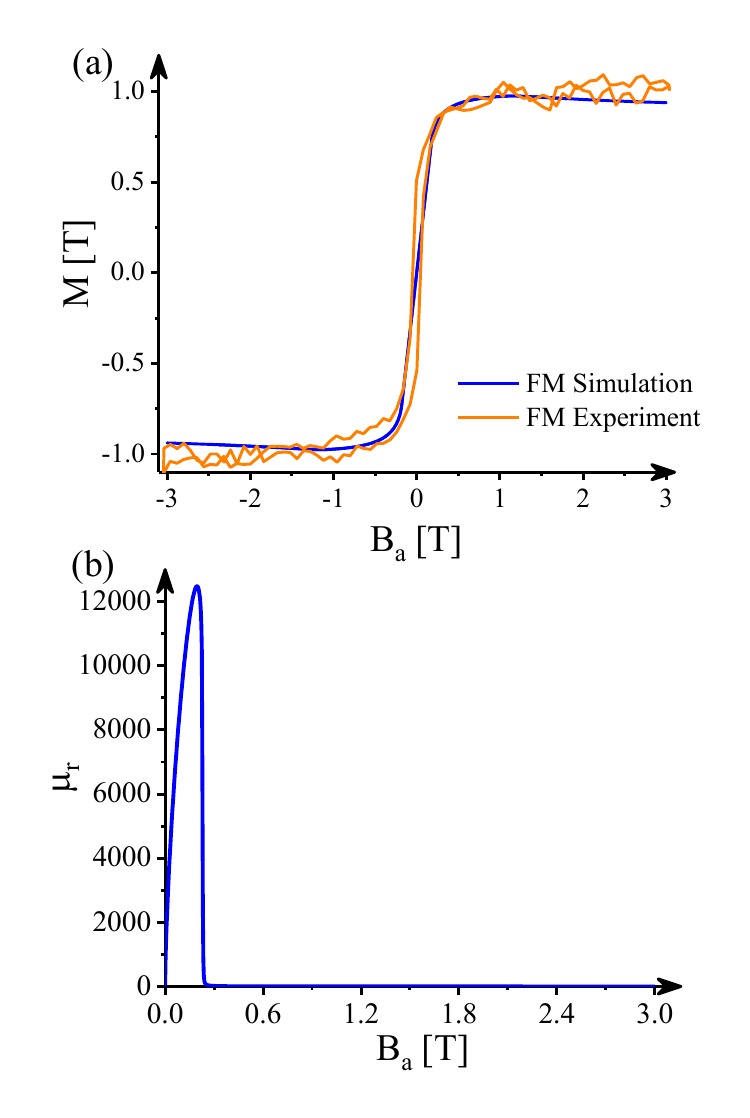}
\caption{(a) The magnetization loop of the FM layer obtained by MPMS measurements. The fitting curve (blue) is in good agreement with the experimental data. (b) The relative permeability variations with applied magnetic field used in the numerical simulations.}
\end{figure}

The maximum field-sweeping rate of MPMS is 0.02 T/s. In this case, we were not able to carry out the experiments with higher field-sweeping rates. Actually, the experimental results are mainly used to verify and confirm the validation of our numerical model and the parameters used in the simulations. Alternatively, to deeply understand the effects of FM layer on the thermomagnetic instabilities of Nb$_3$Sn wire as well as to explore more complicated cases, as schematically shown in Fig. 1(c), we consider a 2D numerical model in the numerical simulations. With the H-formula \cite{M1}, the electromagnetic responses of bare Nb$_3$Sn wire and Nb$_3$Sn/FM wire exposed to a time-varying transverse magnetic field can be obtained by solving the following equations,

\begin{equation}
\begin{aligned}
\mu _{0}\left ( \mu _{r}\left ( {H} \right ) +H _{x}^{2} f\left ({H}\right )    \right )  \frac{\partial H_{x} }{\partial t}+\mu _{0}H_{x}H_{y}f\left ({H}\right )\frac{\partial H_{y} }{\partial t} \\
+ \rho \frac{\partial J_{z}  }{\partial y}=0
\end{aligned}
\end{equation}

\begin{equation}
\begin{aligned}
\mu _{0}\left ( \mu _{r}\left ( {H} \right ) +H _{y}^{2} f\left ( {H}\right )    \right )  \frac{\partial H_{y} }{\partial t}+\mu _{0}H_{x}H_{y}f\left ( {H}\right )\frac{\partial H_{x} }{\partial t} \\
- \rho \frac{\partial J_{z}  }{\partial x}=0
\end{aligned}
\end{equation}

\begin{equation}
\begin{aligned}
\frac{\partial H_{y} }{\partial x} -  \frac{\partial H_{x} }{\partial y} = J_{z}
\end{aligned}
\end{equation}
where $\mu _{0}$ is the vacuum permeability, $\mu _{r}\left ( {H} \right )$ the relative permeability and $\rho$ the resistivity. $H_x(x,y)$ and $H_y(x, y)$ are the components of local magnetic field and thus the total magnetic field ${H}=\sqrt{H_{x}^{2} +H_{y}^{2} }$. $f\left ( {H}\right )=\mathrm{d} \mu _{r}\left ({H} \right )/ \left ({H} \mathrm{d} {H} \right )$.

For the soft 1J22 FM layer, the relationship can be obtained by the magnetization measurements as shown in Fig. 2(a). One can see that a fitting curve agrees with the experimental results quite well. Therefore, similar to Ref \cite{M2}, the following relative permeability is employed in the simulation for the ferromagnetic material,

\begin{equation}
 \mu _{r} \left ( {H}  \right ) =1+1200000\left ( 1-exp\left ( -\left ( {H} /70 \right )  ^{3.2}  \right )   \right){H}^{-0.99}
\end{equation}

The constitutive relationship between current density and electric field is given as
\begin{equation}
\begin{aligned}
\mathbf{E} =\rho \mathbf{J}
\end{aligned}
\end{equation}

The law of resistance in the superconducting region can be written as \cite{M3}
\begin{equation}
\begin{aligned}
\rho _{\scriptscriptstyle SC} = \frac{\rho_{\scriptscriptstyle 0}\rho_{\scriptscriptstyle SC0}  }{\rho _{\scriptscriptstyle 0}+\rho _{\scriptscriptstyle SC0}}
\end{aligned}
\end{equation}

\begin{equation}
\begin{aligned}
\rho _{\scriptscriptstyle SC0} = \frac{E_{c} }{J_{c} } \left ( \frac{\left | J \right | }{\left |J _{c}  \right | }  \right )^{n-1}
\end{aligned}
\end{equation}
where $\rho_{\scriptscriptstyle 0}$ is a constant resistivity, $E_c$ the critical electric field. The critical current density $J_c$ is generally dependent on the temperature and local magnetic field, which can be calculated using the formulae in Appendix. The flux creep exponent $n$ is taken as
\begin{equation}
\begin{aligned}
n= n_{0}\left ( 1-\left (\frac{T}{T_{c}}   \right )^{4}   \right ) \left (1- \frac{B}{B_{c2} }  \right )\frac{T_{c} }{T}
\end{aligned}
\end{equation}
where  $n_0$ is the flux creep exponent at working temperature $T_{0}$, and $T_{c}$ is the critical temperature, $B_{c2}$ is the upper critical magnetic field. The magnetic field $B$ is taken as
\begin{equation}
\begin{aligned}
B = \mu _{0} \mu _{r} H
\end{aligned}
\end{equation}

In order to study the thermomagnetic instability, the electromagnetic equations should be solved by coupling the heating diffusion equations. The heat transport in the system is governed by the equation,

\begin{equation}
\begin{aligned}
c \frac{\partial T}{\partial t} = \bigtriangledown \cdot (\kappa \bigtriangledown T) + \boldsymbol{E} \cdot \boldsymbol{J}
\end{aligned}
\end{equation}
where $\boldsymbol{E} \cdot \boldsymbol{J}$ is the Joule heating source. The heat diffusion equation can be solved by considering the following heat exchange boundary condition
\begin{equation}
\begin{aligned}
-\kappa\left ( \bigtriangledown T\cdot  \mathbf{n} \right )  = h\left ( T-T_{0} \right )
\end{aligned}
\end{equation}
where $c$, $\kappa$, $h$ are the specific heat, thermal conductivity and heat transfer coefficient, respectively. The thermal parameters are assumed to be proportional to $T^3$, i.e., $c=c_{0} \left ( T/T_{0}  \right ) ^3$, $\kappa=\kappa_{0} \left (  T/T_{0} \right ) ^3$, $h=h_{0} \left ( T/T_{0}  \right ) ^3$  \cite{M5}.

\begin{figure*}[!t]
\centering
\includegraphics*[width=0.9\linewidth,angle=0]{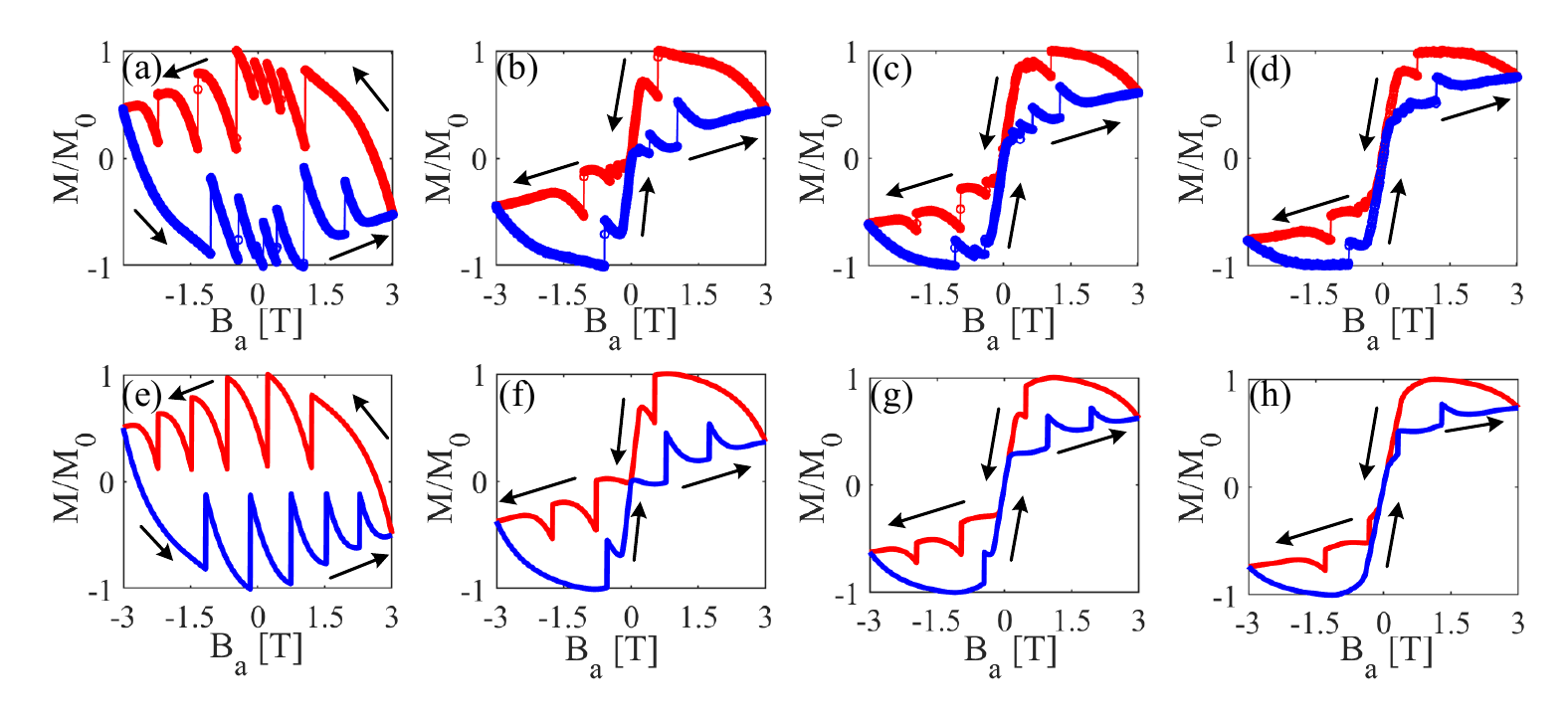}
\caption{(a-d) Experimental magnetization curves of bare Nb$_{3}$Sn wire (sample A) and hybrid Nb$_{3}$Sn wire wrapped by FM layer with thickness of 0.1-0.3 mm (sample B-D) exposed to a transversal magnetic field with a ramping rate of 0.02 T/s by MPMS measurements at 4.2 K. (e-h) Simulated magnetization for sample A-D under the same parameters with experiments. {The red curves denote the magnetization during the process of 3 T to $-3$ T. Similarly, the blue curves denote the magnetization during the process of $-3$ T to 3 T.}}
\end{figure*}

The coupling equations above can be solved by PDE mode in Comsol Multiphysics \cite{M5o,M5t}. In our numerical simulations, the electromagnetic parameters are $J_{c0} = 4\times10^{10}$  $\rm A\cdot m^{-2}$, $T_{0}$=4.2 $\rm K$, $T_{c}$=18.2 $\rm K$, $n_{0}=15$, $E_{c}=1\times 10^{-4}$ $\rm V/m$, $\rho_{air} =1$  $\rm \Omega \cdot  m$, $\rho_{\scriptscriptstyle SC}=2.5\times10^{-7}$ $\rm \Omega \cdot  m$, $\rho_{\scriptscriptstyle Cu}=3\times10^{-10}$ $\rm \Omega \cdot m$, $\rho_{\scriptscriptstyle FM}=9\times10^{-7}$ $\rm \Omega \cdot  m$. After calculation: the temperature of Nb$_3$Sn is the same as that of Cu, the thermal conductivity of Nb$_3$Sn has no effect on the simulation results, and the thermal conductivity of Nb$_3$Sn and Cu is the same. The thermal parameters are $\kappa_{\scriptscriptstyle SC0}=500$ $\rm W/\left (m\cdot K \right )$,  $\kappa_{\scriptscriptstyle Cu0}=500 $ $\rm W/\left (m\cdot K \right )$, $\kappa_{\scriptscriptstyle FM0}= 110$ $\rm W/\left (m\cdot  K \right )$, $c_{\scriptscriptstyle SC0} = 560$ $\rm J/ \left ( K\cdot  m^{3} \right )$, $c_{\scriptscriptstyle Cu0} = 560$ $\rm J/ \left ( K\cdot  m^{3} \right )$, $c_{\scriptscriptstyle FM} =500$ $\rm J/ \left ( K\cdot  m^{3} \right )$ \cite{M6,M7,M8,M9}.

\section{Results and discussions}

%%flux jumps of Samples A-D
\subsection{suppression of flux jumps in Nb$_3$Sn/FM wire exposed to a linear ramping magnetic field loop}

We first start with discussing the flux jumps induced by thermomagnetic instabilities of Nb$_3$Sn/FM wire exposed to a magnetic field loop, i.e., $0$ T $\rightarrow$ 3T $\rightarrow$ -3 T $\rightarrow$ 3 T with a constant field sweeping rate. Fig. 3(a-d) represent the experimental magnetization curves of bare Nb$_{3}$Sn wire, Nb$_3$Sn/FM wire with FM layer thickness of 0.1-0.3 mm, respectively. The sample is in the cold gas helium and the width of the gap between FM layer and Nb$_{3}$Sn wire is about 0.07 mm-0.16 mm during experimental measurements. In order to simulate the partial connection between FM layer and Nb$_3$Sn wire, we assumed that the small gap is filled up with cold gas helium in the numerical model. The thermal parameters of gas helium can be seen in Ref \cite{He1}. Numerical simulations indicated that the temperature is almost the same on the whole cross-section before flux jumps occur. This means that the heat exchange between Nb$_3$Sn wire and FM layer (or Cu layer) is good enough. In this work, we mainly focuse on the flux jumps, i.e., the sharp peaks observed in the magnetization curves. As such, the magnitude of magnetization is not important and the magnetization is normalized by the maximum value M$_{0}$ for the convenient comparison with different samples. By comparing with bare Nb$_{3}$Sn wire (Fig. 3{(}a{)}), we can see a most distinct characteristic of hybrid Nb$_3$Sn/FM wire (Fig. 3(b-d)) that the magnetization curves can only be observed in the first and third quadrants. This is attributed to the strong paramagnetism of 1J22 FM materials and thus the total magnetization M is reversed. {In addition, by comparing the normalized magnetization curves in the four cases, it can be seen that the jump amplitude of Nb$_3$Sn/FM wire is smaller than that of the Nb$_3$Sn wire. Moreover, amplitude of flux jumps is smaller and smaller with increasing thickness of ferromagnetic layer.} Importantly, as shown in Fig. 3(a), frequent flux jumps are observed in bare Nb$_{3}$Sn wire. However, for the Nb$_{3}$Sn wire wrapped by FM layer with thickness of 0.1 mm, the number of the peaks decreases and the flux jumps are suppressed, especially for the decreasing branches of magnetic fields. Moreover, it is surprising to note that the magnetization curves become quite smooth and only 4 flux jumps observed in hybrid Nb$_{3}$Sn/FM wire with FM thickness of 0.3 mm, which indicates that the thick FM layer can significantly suppress the undesired flux jumps triggered in Nb$_{3}$Sn wire.

On one hand, it is apparent that suppression of flux jumps by wrapping FM layer should first be attributed to the thermal effect because it increases the total heating capacity, which definitely enhances the thermomagnetic stability of Nb$_{3}$Sn wire exposed to a ramping magnetic fields. The main origin of the idea is based on the well-known fact that the flux jumps are thermomagnetic coupling problems. As a matter of fact, it had been demonstrated that power-doping with high specific heating capacity in Nb$_{3}$Sn wire, such as Gd$_2$O$_3$, is an effective way to suppress the flux jumps \cite{B10} . On the other hand, one could possibly argue that, besides the indirect effect, the FM layer should also have an extra impact on the flux jumps due to its electromagnetic effect. The main argument behind this belief leans on the fact that FM layer is a typical paramagnetic material and thus can directly affect the electromagnetic responses of the Nb$_{3}$Sn wire. In order to clarify the underlying physics, we performed numerical simulations for the bare Nb$_{3}$Sn wire and hybrid Nb$_{3}$Sn/FM wire exposed to the same magnetic field loop with the aforementioned magnetization experiments. As shown in Fig. 3(e-h), the simulated magnetization of bare Nb$_{3}$Sn wire and Nb$_{3}$Sn/FM wire can reproduce the experimental results quite well. For Nb$_{3}$Sn/FM system, both experimental and simulated results demonstrate that the flux jumps are triggered at the increasing branches ($0 \rightarrow \pm 3$) T more frequently than at the decreasing branches ($\pm 3 \rightarrow 0$) T.

%($n_s$ as discussed in Ref. \cite{drag})
\begin{figure}[!t]
\centering
\includegraphics*[width=1.0\linewidth,angle=0]{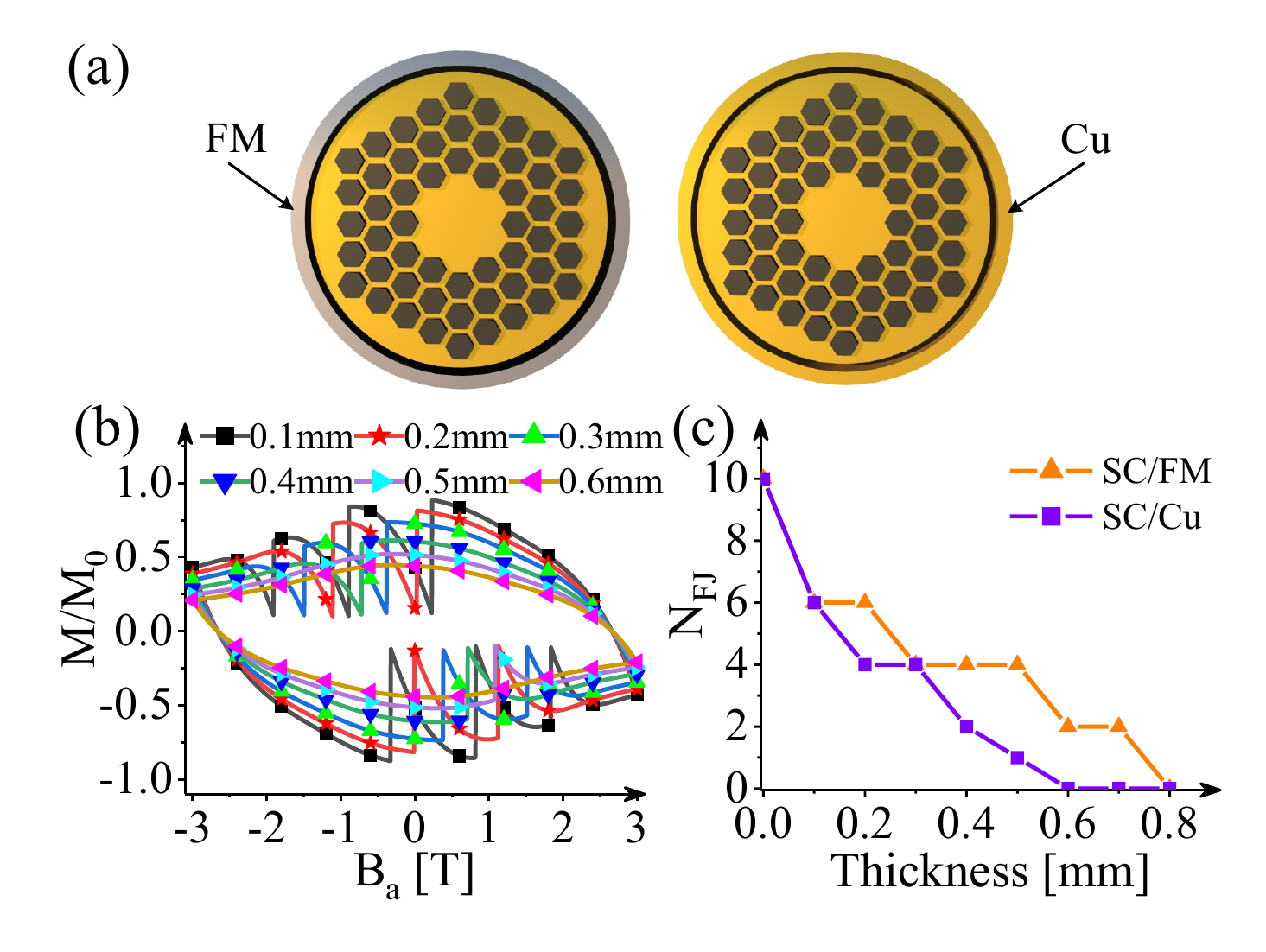}
\caption{(a) Schematic diagrams of two same-sized Nb$_{3}$Sn/FM wire (left) and Nb$_{3}$Sn/Cu wire (right) used in the numerical simulations. (b) The magnetization curves of Nb$_{3}$Sn/Cu wire with Cu thickness of 0.1-0.6 mm. (c) The number of flux jumps {($\rm N_{\rm\scriptscriptstyle FJ}$)} as a function of the thickness of Cu and FM layer.}
\end{figure}

To further explore the thermal effect and electromagnetic effect of FM layer on suppression of flux jumps, as illustrated in Fig. 4(a), we calculate the magnetization of Nb$_{3}$Sn wire with Cu layer which has the same structure size with Nb$_{3}$Sn/FM wire. Although the Cu can lead to modifications of the current streamlines over the superconducting wire after quench, it is worth noting that, as shown in Fig. 3, the magnetization of Nb$_{3}$Sn wire does not decrease to zero during the flux jumps and thus Nb$_{3}$Sn wire does not quench yet. So the {induced current} still flows in the Nb$_{3}$Sn filaments rather than in cooper during the flux jumps and this is also verified by our numerical simulations. In this case, the Cu layer only exhibits thermal effect on the flux jumps while its electromagnetic effect can be neglected. Fig. 4(b) shows the magnetization of Nb$_{3}$Sn/Cu wire with Cu layer thickness of {0.1-0.6mm}. {It can be seen that the number and amplitude of the flux jumps of the Nb$_3$Sn/Cu wire decrease with increasing the thickness of the Cu layer. Particularly, the flux jumps of the hybrid structure can be completely suppressed when the thickness of Cu layer is increased up to 0.6 mm.} Fig. 4(c) indicates that the flux jumps are less likely to occur when both increasing the thickness of Cu layer and FM layer. Additionally, the Cu layer exhibits a slightly more powerful suppression of flux jumps due to its larger specific heating capacity of Cu as mentioned in section \uppercase\expandafter{\romannumeral2}. To conclude, the suppression of flux jumps by FM layer is mainly attributed to its thermal effect, while its electromagnetic effect is ineffective.

\begin{figure}
\centering
\includegraphics*[width=1.0\linewidth,angle=0]{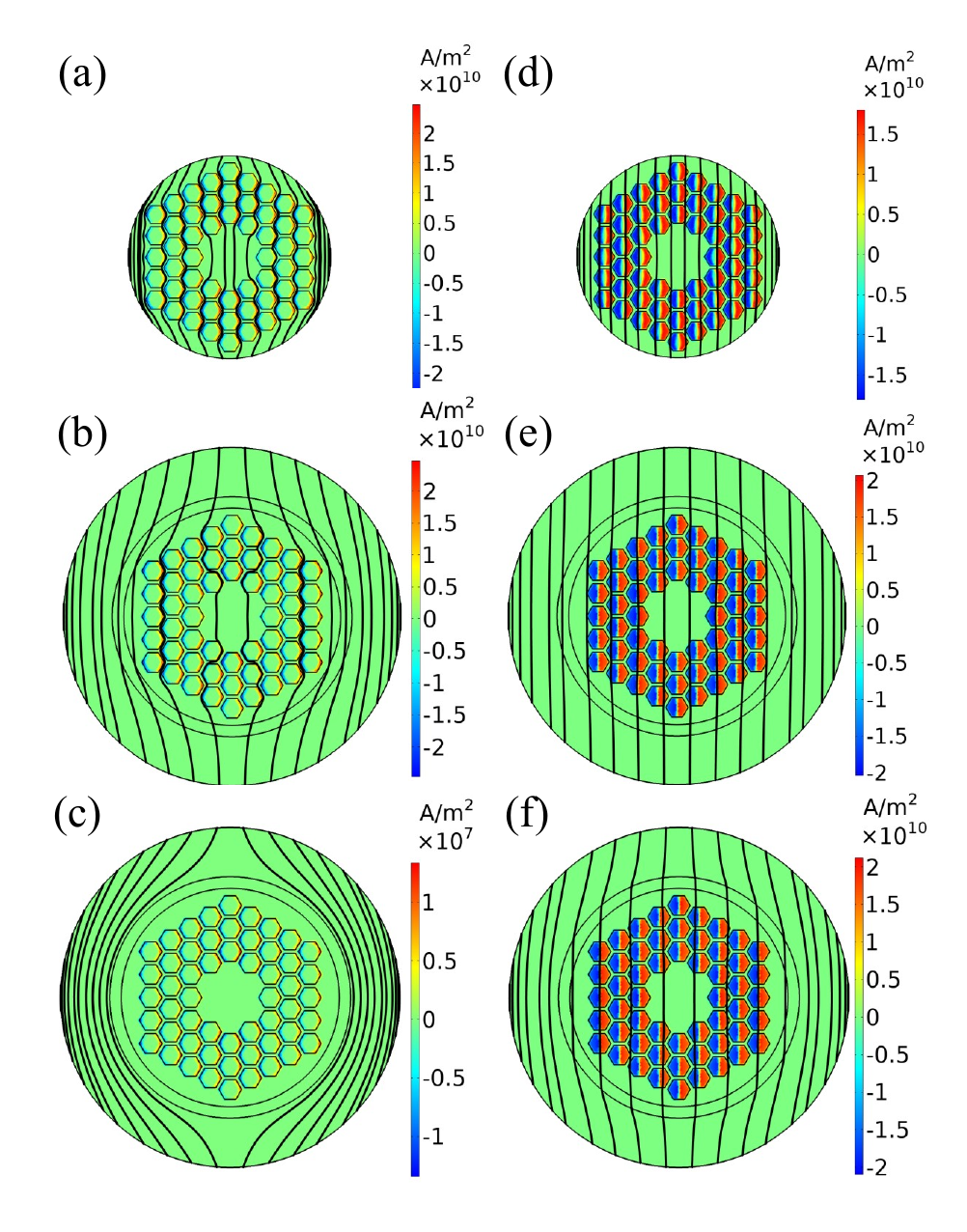}
\caption{The simulated magnetic flux lines and {the induced} current density distributions over bare Nb$_{3}$Sn wire (a, d), Nb$_{3}$Sn/Cu wire (b, e) and Nb$_{3}$Sn/FM wire (c, f) exposed to $B_{a}$= 0.2 T (left panels) and 3.0 T (right panels).}
\end{figure}

In order to deeply understand the phenomenon that seems to be against the common sense, Fig. 5 depicts the magnetic flux lines and {the induced} current density distributions over the bare Nb$_{3}$Sn wire, Nb$_{3}$Sn/FM wire and Nb$_{3}$Sn/Cu wire exposed to a lower and higher magnetic fields, respectively. In the following text, the thicknesses of FM layer and Cu layer are both fixed to be 0.3 mm in all numerical simulations unless otherwise stated. For the case of lower magnetic field, it can be found that the induced current density in bare Nb$_{3}$Sn wire and Nb$_{3}$Sn/Cu wire are nearly the same, which are much higher than that in Nb$_{3}$Sn/FM wire. Moreover, most of magnetic flux lines pass through the FM layer, which demonstrates a significant magnetic shielding effect. However, the Cu layer has no impact on the distributions of magnetic flux lines. In contrast to the low magnetic field, the impact of FM layer on the electromagnetic responses of Nb$_{3}$Sn becomes much weaker in the case of higher magnetic fields. As shown in Fig. 2, the experimental measurements show that the magnetization of 1J22 FM material is saturated and its relative permeability decreases rapidly with increasing magnetic field when $B_a >$ 0.25 T. As a consequence, in the case of small ramping rate $\dot{B_{a}}$, the ferromagnetic layer has nearly no extra advantage over increasing the ratio of cooper to Nb$_{3}$Sn due to the rapid decrease of relative permeability of ferromagnetic material at higher magnetic fields.

\begin{figure}
\centering
\includegraphics*[width=0.9\linewidth,angle=0]{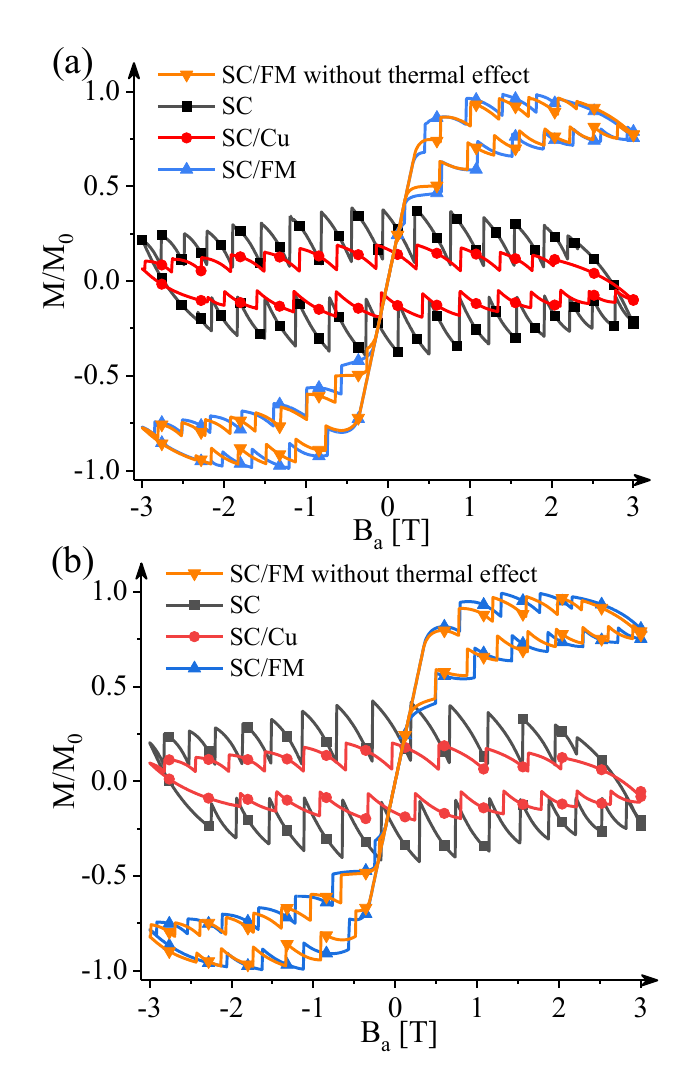}
\caption{(a-b) The simulated magnetization of bare Nb$_{3}$Sn wire, Nb$_{3}$Sn/Cu wire and Nb$_{3}$Sn/FM wire exposed to a magnetic field with a higher field-sweeping rate of $\dot{B_{a}}$=0.1 T/s for $h=4$ and 8 {$\rm W/\left ( K\cdot  m^{2} \right )$}, respectively. The thickness of Cu and FM layers are 0.3 mm. For comparison, both cases of the Nb$_{3}$Sn/FM wire with/without thermal effect are presented.}
\end{figure}

\begin{figure}
\centering
\includegraphics*[width=0.9\linewidth,angle=0]{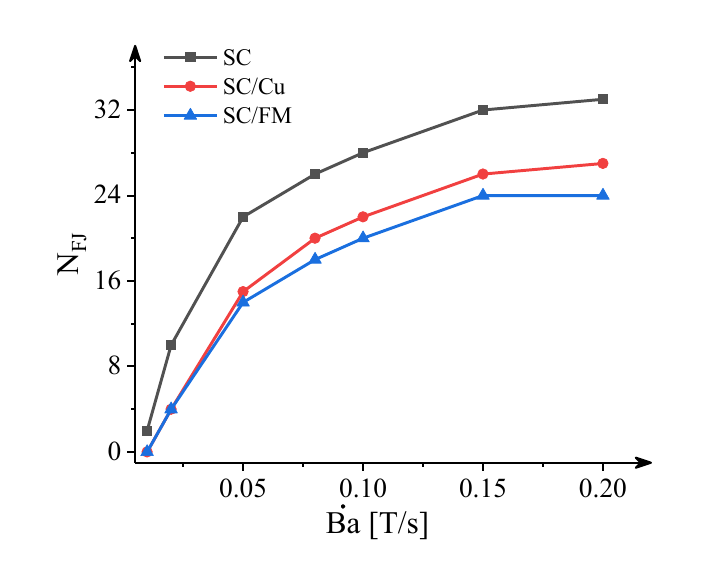}
\caption{The number of flux jumps in bare Nb$_{3}$Sn wire, Nb$_{3}$Sn/Cu wire and Nb$_{3}$Sn/FM wire triggered during one applied magnetic field loop with different ramping rate.}
\end{figure}

Now we explore the case of three kinds of superconducting wires exposed to a magnetic field loop with a higher field-sweeping rate. Fig. 6 shows simulated magnetization of bare Nb$_{3}$Sn wire, Nb$_{3}$Sn/Cu wire and Nb$_{3}$Sn/FM wire with $\dot{B_{a}}$=0.1 T/s. Counting the peaks in magnetization curves, one can see that the flux jumps of Nb$_{3}$Sn/FM wire is less than that of Nb$_{3}$Sn/Cu wire, which indicates that the FM layer has better suppression of flux jumps than Cu layer for higher field-ramping rate. It seems that the electromagnetic effect of FM layer on Nb$_{3}$Sn wire is enhanced for the cases of higher field-ramping rate. Note that it is not an accident because the similar results can also be observed by varying the heating transfer coefficient as shown in Fig. 6(b). In order to extract the electromagnetic effect of FM layer on the flux jumps, we remove the heating effect in the numerical simulations. As shown in Fig. 7, the mere electromagnetic effect of FM layer can indeed suppress the flux jumps partially. One may suppose that the suppression of flux by ferromagnetic layer depends on the field-ramping rate. Fig. 7 represents the number of flux jumps in the bare Nb$_{3}$Sn wire, Nb$_{3}$Sn/Cu wire and Nb$_{3}$Sn/FM wire with increasing the field-ramping rate of applied magnetic field $\dot{B_{a}}$. The results demonstrate that flux jumps are more likely triggered in the three kinds of Nb$_{3}$Sn wires with increasing $\dot{B_{a}}$. It is interesting to note that the flux jumps triggered in Nb$_{3}$Sn/FM wire is less than the Nb$_{3}$Sn/Cu wire in the case of higher field-ramping rate. This demonstrates that advantage of FM layer over the Cu layer becomes more and more apparent with increasing $\dot{B_{a}}$ although the FM layer and Cu layer have nearly the same suppression of flux jumps for small $\dot{B_{a}}$.

\begin{figure}
\centering
\includegraphics*[width=1.0\linewidth,angle=0]{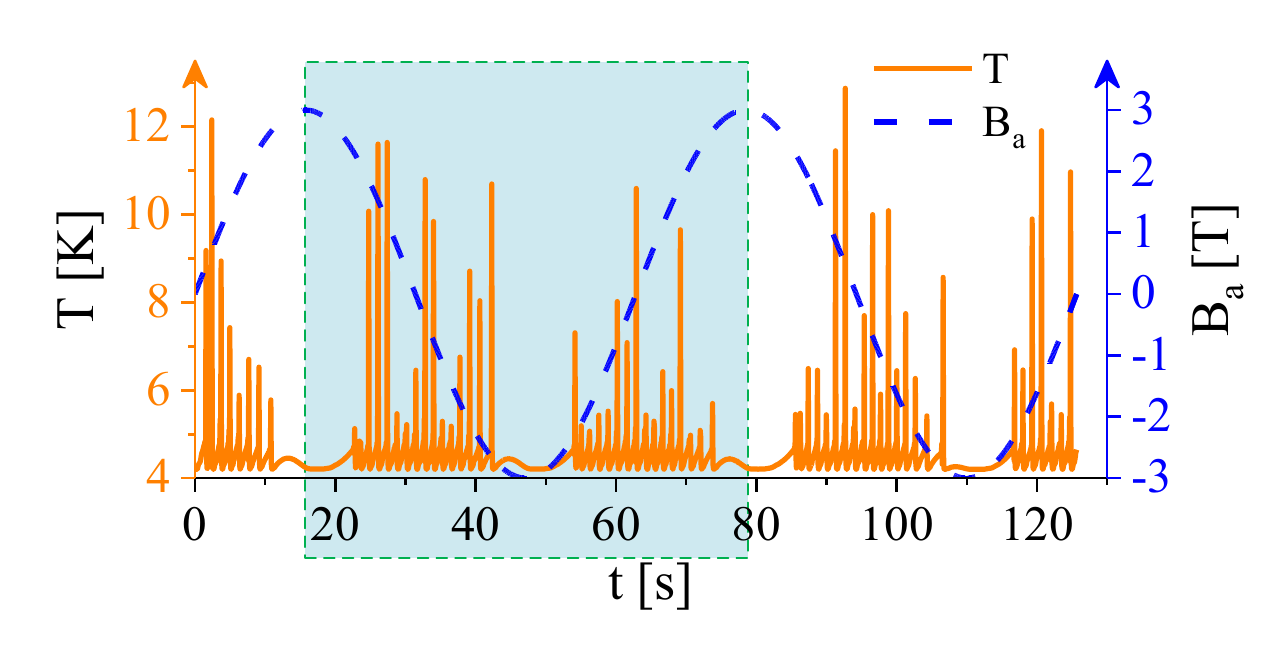}
\caption{The variation of AC magnetic field with time used in the numerical simulations and the time evolutions of simulated temperature of Nb$_{3}$Sn wire.}
\end{figure}

\subsection{Suppression of flux jumps in Nb$_{3}$Sn/FM wire exposed to an AC magnetic field}

In this subsection, we further explore the suppression of flux jumps in Nb$_{3}$Sn wire exposed to an AC magnetic field by FM layer. {Due to the magnetization loss of ferromagnetic materials under AC magnetic fields, local heating may occur. However, we choose soft ferromagnetic material. Unlike hard ferromagnetic material, the soft ferromagnetic material has very small magnetization loop (see experimental magnetization of only ferromagnetic layer in Fig. 2). Furthermore, we also calculated the magnetization loss of ferromagnetic layer with the numerical simulations. As shown in Fig. A1 in the Appendix, the magnetization loss of ferromagnetic layer is very small, which can be neglected.} Unlike the cases of a constant field ramping rate in subsection A, the main characteristic of AC magnetic field is that the field ramping rate always varies with time, and it also depends on the present magnetic field. As shown in Fig. 8, we consider an AC magnetic field with sinusoidal variations, i.e., $B_a(t)=B\rm_{a0} sin$$(2\pi {f}t)$. The frequency {$f$} and the amplitude $B_{a0}$ of the AC magnetic field can be tunable in the numerical simulations. Although the signals of flux jumps are not strictly periodic with AC magnetic field, it can be found that the number of simulated flux jumps are nearly the same in each period except the initial increasing branch. Therefore, we count the flux jumps within a period of AC magnetic field after the initial increasing branch (see colored box in Fig. 8).

\begin{figure}
\centering
\includegraphics*[width=1.0\linewidth,angle=0]{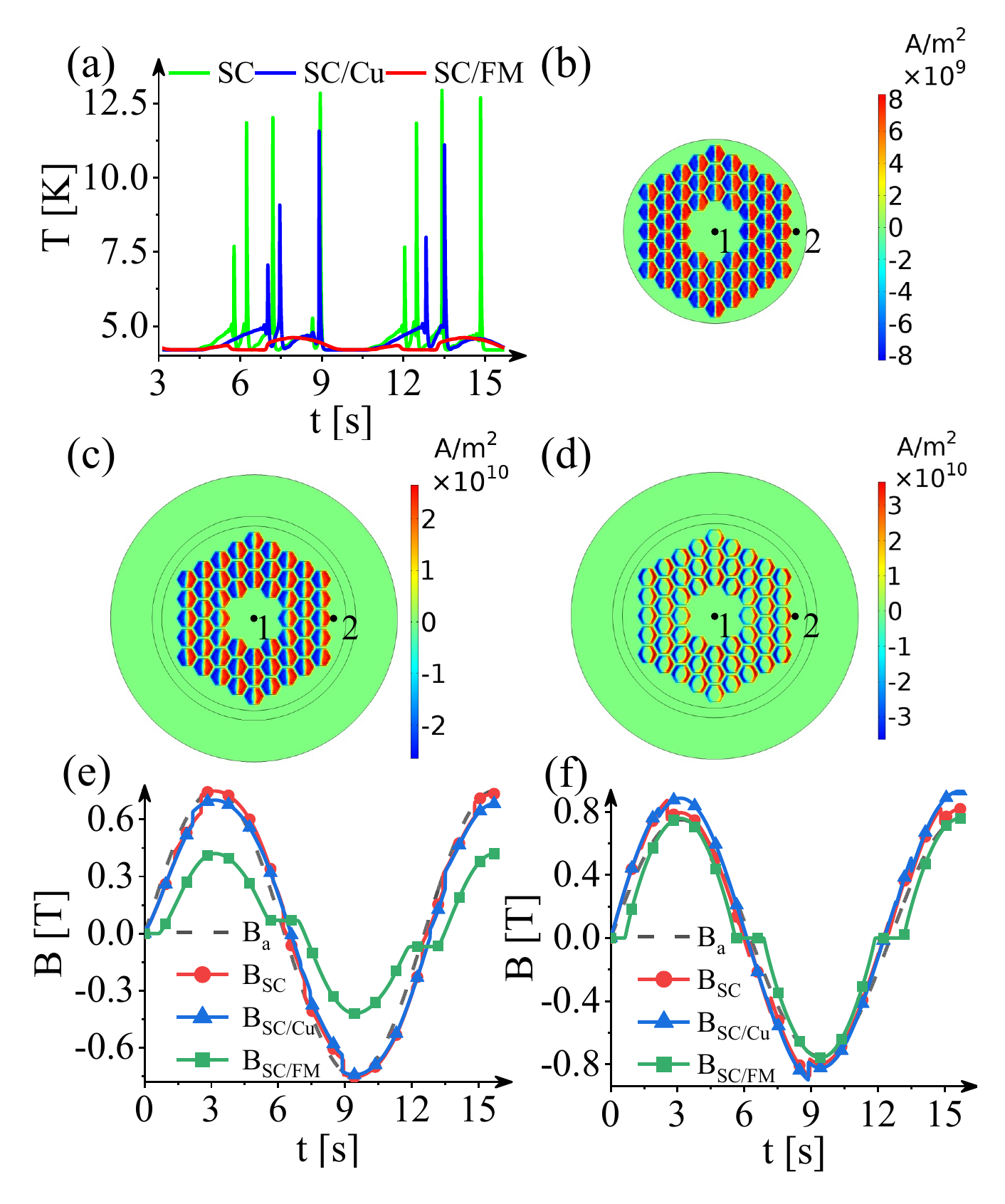}
\caption{(a) The time evolution of temperature in bare Nb$_{3}$Sn wire, Nb$_{3}$Sn/Cu wire and Nb$_{3}$Sn/FM wire exposed to an AC magnetic field with frequency {$f=0.5$} Hz and $B_{a0}=0.75$ T. (b-d) The current density distributions over the three kinds of wires when $B_{a}(t)=0.75$ T. (e-f) The local magnetic field evolutions with time at point 1 and 2 indicated in (b-d). The dashed black curves are the variations of AC magnetic field with time.}
\end{figure}

We first explore a simple case of AC magnetic field by fixing the frequency and amplitude. Fig. 9(a) shows the time evolutions of temperature in bare Nb$_{3}$Sn wire, Nb$_{3}$Sn/Cu wire and Nb$_{3}$Sn/FM wire exposed to an AC magnetic field with {$f=0.5$} Hz and $B_{a0}=0.75$ T. The averaged ramping rate of magnetic field is about 1.5 T/s, which is much greater than the field-sweeping rates investigated above. By noting that flux jumps triggered in Nb$_{3}$Sn/Cu wire is less than that in bare Nb$_{3}$Sn wire, one can see that the Cu layer can partially suppress the flux jumps, which is similar to the cases in subsection A. By contrast, it is surprising that the FM layer can completely eliminate the flux jumps. To understand the difference of thermomagnetic instabilities in three kinds of Nb$_{3}$Sn wires exposed to AC magnetic field, Fig. 9(b-d) represents the current density distributions over bare Nb$_{3}$Sn wire, Nb$_{3}$Sn/Cu wire and Nb$_{3}$Sn/FM wire when $B_a(t)=0.75$ T, respectively. It can be observed that both bare Nb$_{3}$Sn wire and Nb$_{3}$Sn/Cu wire are completely penetrated by magnetic flux. However, the magnetic flux just partially penetrates the Nb$_{3}$Sn/FM wire. The maximum current density of Nb$_{3}$Sn/FM wire is greater than the bare Nb$_{3}$Sn wire and Nb$_{3}$Sn/Cu wire because flux jumps occur before $B_a(t)=0.75$ T in the latter two cases. Additionally, one may expect to find more evidence from the local magnetic field to interpret the suppression of flux jumps by FM layer. Fig. 9(e-f) shows the time evolutions of two representative local magnetic field at point 1 and 2 indicated in Fig. 9(b-d). It is evident that the local magnetic field $B_{\rm\scriptscriptstyle SC/FM}$ at point 1 and 2 is always much less than $B_{\rm \scriptscriptstyle SC}$ and $B_{\rm\scriptscriptstyle SC/Cu}$. Therefore, the FM layer decreases not only the local magnetic fields, but also the field ramping rate.

\begin{figure}
\centering
\includegraphics*[width=1.0\linewidth,angle=0]{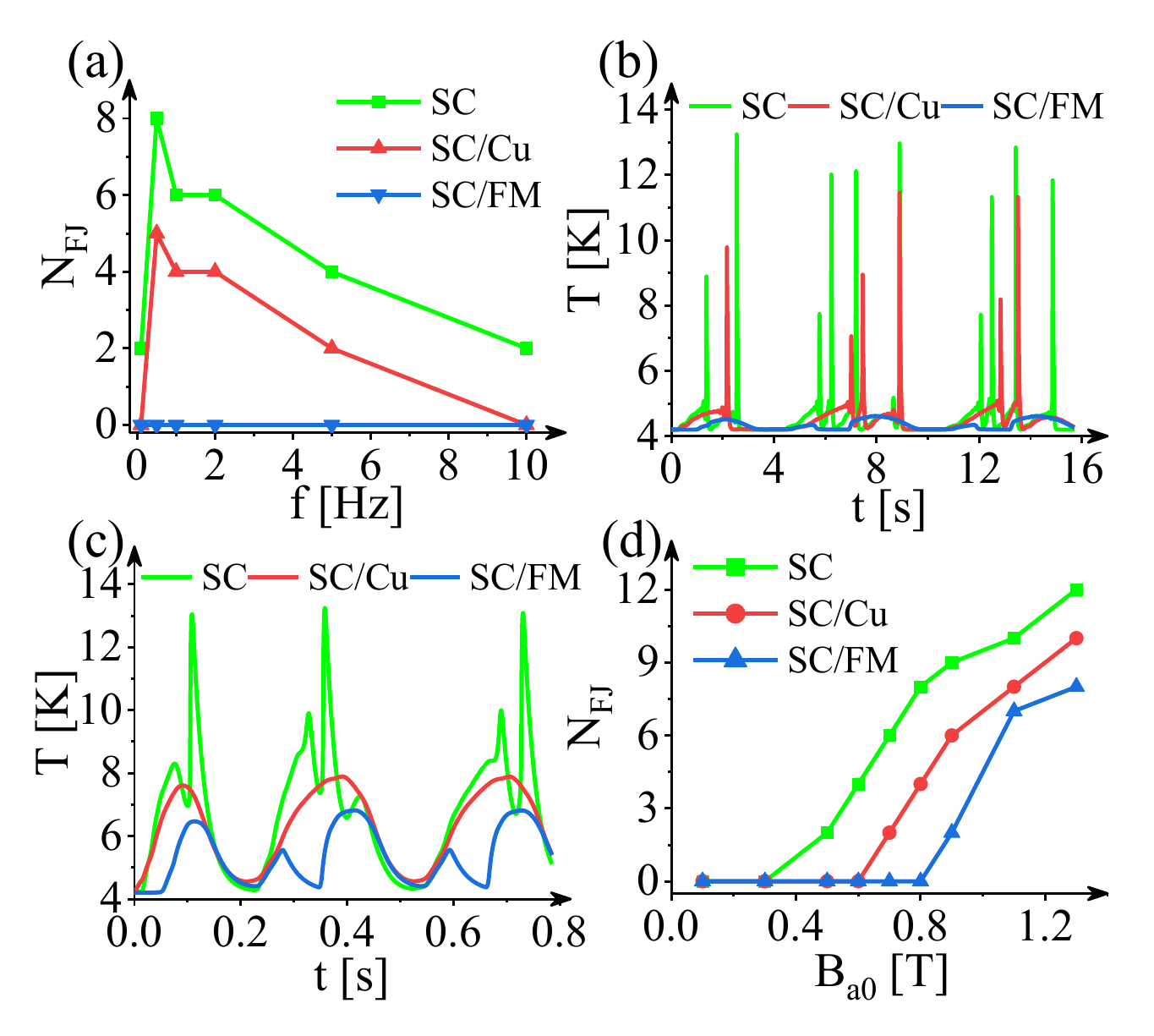}
\caption{(a) The number of flux jumps as a function of frequency $\rm f$ in bare SC wire, SC/Cu wire and SC/FM wire exposed to AC magnetic field by fixing $B_{a0}=0.75$ T. (b-c) The time evolutions of temperature in three kinds of wires for {$f=0.5$} Hz and 10 Hz, respectively. (d)The number of flux jumps as a function of amplitude of ac magnetic field by fixing {$f=0.5$} Hz.}
\end{figure}

For $B_{a0}=0.75$ T, Fig. 10(a) indicates that the flux jumps are completely eliminated by the FM layer for all frequencies of AC magnetic field, while the flux jumps are observed in bare Nb$_{3}$Sn wire and Nb$_{3}$Sn/Cu wire. Furthermore, the number of flux jumps triggered in both bare Nb$_{3}$Sn wire and Nb$_{3}$Sn/Cu wire is not a monotonous function with frequency of AC magnetic field. This can be interpreted by the temperature. Fig. 10(b-c) shows the time evolutions of temperature in three kinds of wires for {$f=0.5$} Hz and {$ f=10$} Hz, respectively. By taking the Nb$_{3}$Sn/Cu wire as an example, the temperature varies in the range of 4.2-7.6 K for {$ f=10$} Hz (see Fig. 10{(}c{)}), while the maximum temperature is less than 5 K before triggering flux jumps for {$ f=0.5$} Hz (see Fig. 10{(}b{)}). As studied in previous work (see Refs. \cite{S1}), the pinning effect on magnetic vortices is weakened and the critical current density is reduced at higher temperature. This leads to the magnetic flux penetrates into the superconductor easier and the magnetic pressure is not sufficiently large to triggered the flux jumps. As a consequence, the smooth flux penetrations are statistically more likely to occur and the thermomagnetic instabilities are less likely triggered in this case. Due to the field-shielding effect of FM layer, one can observe that the temperature of Nb$_{3}$Sn/FM wire is much smaller than the other two kinds of wires. Although the minimum temperature is about 4.2 K in Nb$_{3}$Sn/FM wire for {$ f=10$} Hz, the ramping rate of magnetic field is very small and flux jumps cannot be triggered at that time.

For the case of fixing frequency of AC magnetic field {$ f=0.5$} Hz, Fig. 10(d) shows that the flux jumps are triggered in bare Nb$_{3}$Sn wire when the AC magnetic field amplitude $B_{a0}>0.3$ T. By contrast, the threshold amplitude of AC magnetic field can be increased to 0.6 T and 0.8 T in Nb$_{3}$Sn/Cu wire and Nb$_{3}$Sn/FM wire, respectively. On one hand, this again demonstrates that the Cu layer and FM layer can enhance the thermomagnetic stabilities of Nb$_{3}$Sn wire exposed to AC magnetic field. On the other hand, by combining with the fact that the specific heating capacity of cooper is a bit greater than that of ferromagnetic material 1J22, it also provides the evidence for that the mechanism of better suppression of flux jumps by FM layer than the Cu layer is attributed to both thermal effect and electromagnetic effect.

\begin{figure}
\centering
\includegraphics*[width=0.9\linewidth,angle=0]{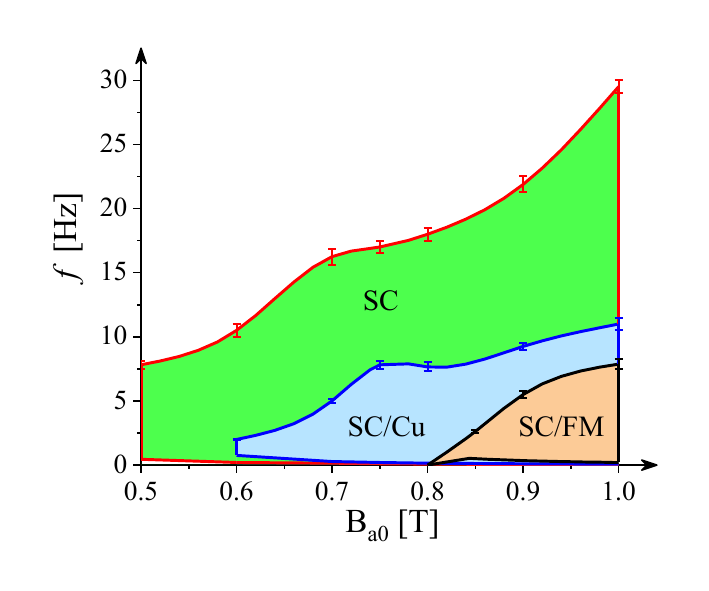}
\caption{The phase diagram of flux jumps in the plane {$f$-$B_{a0}$} for bare Nb$_{3}$Sn wire, Nb$_{3}$Sn/Cu wire and Nb$_{3}$Sn/FM wire. The different colored curves represent the threshold frequencies as functions of amplitude of AC magnetic field.}
\end{figure}

In the last, we explore more complicated situations by varying both amplitude and frequency of AC magnetic fields. Fig. 11 shows the phase diagram of thermomagnetic instabilities in plane {$f$-$B_{a0}$} and the different colors denote the regions of flux jumps triggered in bare Nb$_{3}$Sn wire, Nb$_{3}$Sn/Cu wire and Nb$_{3}$Sn/FM wire, respectively. Both lower and upper threshold frequencies can be observed in the phase diagrams, which indicates that the flux jumps can only be observed in AC magnetic fields with a specific range of frequencies by fixing the amplitude. Additionally, the threshold frequencies for all three kinds of wires depends on the amplitude of AC magnetic fields. Most importantly, the region of flux jumps for Nb$_{3}$Sn/FM wire is much smaller than other two kinds of wires. As a consequence, the hybrid Nb$_{3}$Sn/FM wire indeed has more advantage to improve the thermomagnetic stability than bare Nb$_{3}$Sn wire and Nb$_{3}$Sn/Cu wire when it is used in AC magnetic fields.

\section{Conclusions}

In summary, we experimentally and theoretically explore suppressing the undesirable low-field instabilities of high-$J_c$ Nb$_{3}$Sn conductors by FM materials. We fabricate two kinds of sample of bare Nb$_{3}$Sn and Nb$_{3}$Sn/FM hybrid wires and implement magnetization measurements. Compared with bare Nb$_{3}$Sn wire, the experiments indicate that the flux jumps decrease significantly in Nb$_{3}$Sn/FM hybrid wires. We find that the thicker the FM layer is, the less flux jumps are observed. In order to reveal the underlying mechanism, we also implement numerical simulations. The validation of the numerical simulation are verified by reproducing the experiments quite well. We further numerically study two cases of Nb$_{3}$Sn/FM hybrid wires exposed to magnetic field loop with constant ramping rate and AC magnetic field with time-varying ramping rate, respectively. {Even though the effective critical current density of Nb$_3$Sn wire is decreased by the outer layer because the cross-section of the wire is increased. As shown in Fig. A2 in the Appendix, simulation results indicate that when the critical current of the bare Nb$_{3}$Sn wire is reduced to the same as the FM/Nb$_{3}$Sn wire equivalent critical current density, the FM layer is still effective in suppressing flux hopping due to the additional electromagnetic effect.} The main results are concluded as follows.

(a) When the field-sweeping rate is small, we find that the flux jumps of Nb$_{3}$Sn suppressed by FM layer are mainly attributed to the thermal effect, i.e., increasing the total heat capacity. As for higher field-sweeping rates, less flux jumps are observed in Nb$_{3}$Sn/FM hybrid wire than in Nb$_{3}$Sn/Cu wire, which indicates that both thermal and eletromagnetic effects of FM materials are crucial to suppress the flux jumps in this case. 

(b) For Nb$_{3}$Sn/FM hybrid wires exposed to AC magnetic field, it is found that the flux jumps can be suppressed completely, which depends on the amplitude and frequency of AC magnetic field. The local magnetic field inside Nb$_{3}$Sn/FM hybrid wire and the temperature rise is significantly smaller than that in bare Nb$_{3}$Sn wire and Nb$_{3}$Sn/Cu wire. We further build up a phase diagram in the plane {$f$-$B_{a0}$} for bare Nb$_{3}$Sn wire, Nb$_{3}$Sn/Cu wire and Nb$_{3}$Sn/FM wire, respectively. The region of flux jumps in Nb$_{3}$Sn/FM wire is much smaller than other two kinds of wires, which demonstrates that the FM materials exhibit much better suppression of low-field instabilities than merely increasing the heat capacity.

\vspace{6ex}
\noindent
\textbf{Acknowledgement}
\vspace{1ex}

The authors acknowledge support by the the National Natural Science Foundation of China (Grant Nos. 12372210, 11972298, 12174242).

\vspace{6ex}
\noindent
\textbf{Appendix}
\vspace{1ex}

The critical current density $J_c$ is generally dependent on the temperature and local magnetic field, which can be expressed as \cite{P1,P2}
\setcounter{equation}{0}
\renewcommand{\theequation}{A\arabic{equation}}
\begin{equation}
\begin{aligned}
J_{c}\left ( B,T \right )  =  J_{c}\left ( B,T_{0}  \right )\frac{\left ( 1+\left ( \frac{T}{T_{c}}  \right )^{2}\right )^{-0.5} \left ( 1-\left ( \frac{T}{T_{c}}  \right )^{2}\right )^{2.5}  }{\left ( 1+\left ( \frac{T_{0}}{T_{c}}  \right )^{2}\right )^{-0.5} \left ( 1-\left ( \frac{T_{0}}{T_{c}}  \right )^{2}\right )^{2.5}}
\end{aligned}
\end{equation}
with
\begin{equation}
\begin{aligned}
J_{c}\left ( B,T_{0}  \right )=\frac{b}{A} \left ( \frac{B}{B_{c2} } \right )  ^{-p} \left ( 1- \frac{B}{B_{c2} }\right )^{q}
\end{aligned}
\end{equation}
where $a=25$, $b=7731.5$, $p=0.4$, $q=3.12$, $A=6.65\times10^{-7}$.

%%%

\renewcommand{\thetable}{A\arabic{table}}

\begin{table}
  \begin{center}
    \caption{Sample data}
    \label{tab:table1}
    \begin{tabular}{c|c|c|c|c} 
	  \hline
	  Sample & FM thickness & Diameter & Subelements & Gap \\
      \hline
	  SC    & \textbackslash{}  & 1.3mm   & 54          & \textbackslash{} \\
	  \hline
	  		& 0.1mm              & 1.82mm   & 54          & 0.16mm           \\
	  \cline{2-5}
      SC/FM	& 0.2mm              & 1.92mm   & 54          & 0.11mm           \\
	  \cline{2-5}
      		& 0.3mm              & 2.04mm   & 54          & 0.07mm           \\
	  \hline
    \end{tabular}
  \end{center}
\end{table}

\setcounter{figure}{0}
\renewcommand{\thefigure}{A\arabic{figure}}
\begin{figure}
\centering
\includegraphics*[width=0.9\linewidth,angle=0]{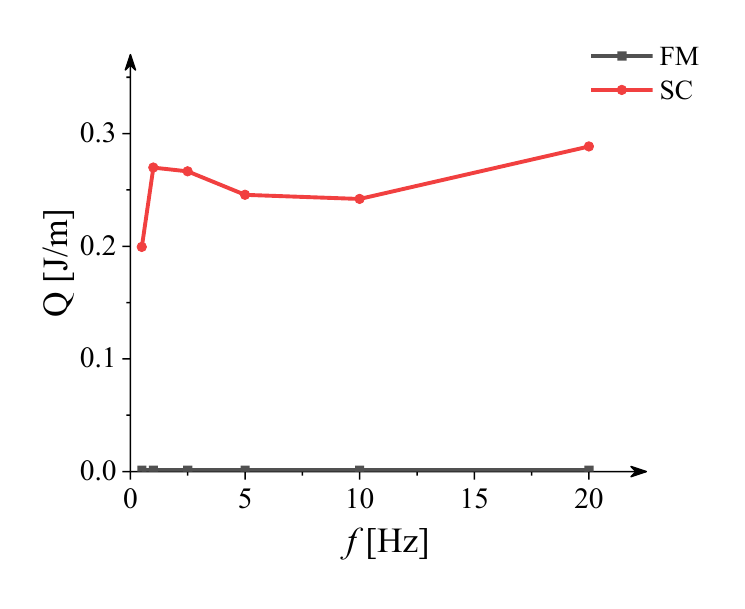}
\caption{Simulated magnetization loss of ferromagnetic layer and Joule heating of superconducting wire.}
\end{figure}

\begin{figure}
\centering
\includegraphics*[width=0.9\linewidth,angle=0]{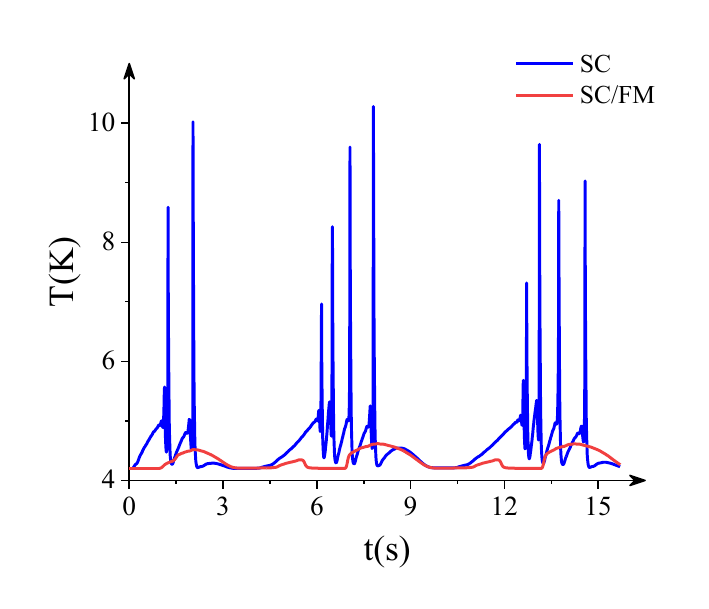}
\caption{Temperature curves of bare SC wire and SC/FM wire with the same equivalent critical current.}
\end{figure}


\end{document}